\documentclass[cpp,a4paper,fleqn%
]{w-art}
\usepackage{times,cite,w-thm}

\theoremstyle{plain}

\theoremstyle{definition}

\usepackage[]{graphicx}
\usepackage{textcomp}
\begin{document}
\DOIsuffix{theDOIsuffix}
\Volume{46}
\Month{01}
\Year{2007}
\pagespan{1}{}
\Receiveddate{XXXX}
\Reviseddate{XXXX}
\Accepteddate{XXXX}
\Dateposted{XXXX}



\title[Clusters in nanocomposites]{Diffusion and growth of metal clusters in nanocomposites: \\a Kinetic Monte Carlo study\footnote{Based on a lecture at the second Summer Institute ``Complex Plasmas'', Greifswald, August 2010.}}


\author[L. Rosenthal et al.]{L. Rosenthal\inst{1}}
\address[\inst{1}]{Institut f\"ur Theoretische Physik und Astrophysik, Christian-Albrechts-Universit\"at Kiel, Leibnizstrasse 15, 24098 Kiel, Germany}
\author[]{A. Filinov\inst{1}}
\author[]{M. Bonitz\inst{1}\footnote{Corresponding author\quad E-mail:~\textsf{bonitz@physik.uni-kiel.de},
            Phone: +49\,431\,880\,4122.}}
\address[\inst{2}]{Lehrstuhl f\"ur Materialverbunde, Technische Fakult\"at, Christian-Albrechts-Universit\"at Kiel, Kaiserstrasse 2, 24143 Kiel, Germany}
\author[]{V. Zaporojtchenko\inst{2}}
\author[]{F. Faupel\inst{2}}

\newcommand{\printwidthof}[2][pt]{\setlength{\dimen1}{\widthof{#2}}\uselengthunit{#1}\printlength{\dimen1}}
\newcommand{\printheightof}[2][pt]{\setlength{\dimen1}{\heightof{#2}}\uselengthunit{#1}\printlength{\dimen1}} 


\begin{abstract}
Nobel metals that are deposited on a polymer surface exhibit surface diffusion and diffusion into the bulk.
At the same time the metal atoms tend to form clusters because their cohesive energy is about two orders of magnitude higher than the cohesive energy of polymers. To selfconsistently 
simulate these coupled processes, we present in this paper a Kinetic Monte Carlo approach. Using a simple model with diffusion coefficients taken as input parameters allows us to perform a systematic study of the behavior of a large ensemble of metal atoms on a polymer surface eventually leading to polymer nanocomposites. Special emphasis is placed on the cluster growth, cluster size distribution and the penetration of clusters into the substrate. We also study the influence of surface defects and analyze how the  
properties of the resulting material can be controlled by variation of the deposition rate.
\end{abstract}
\maketitle         

\section{Introduction}
Metal-polymer nanocomposites are finding growing application in many areas of science and technology because the combination of the diverse physical features of their constituents allow for the production of materials with interesting novel properties. Examples of these properties include a broad range of electronic conductivity, while maintaining the high mechanical flexibility of the substrate, e.g.\cite{deposition_by_plasmas}, special optical absorption and reflectivity and even anti-microbial properties\cite{bib2}. 
A particularly promising route to the production of nanocomposites is sputtering of metal atoms and polymer in a plasma environment, e.g., using magnetrons\cite{magnetron}, as they allow to vary the deposition parameters in a broad range. 

For efficient manufacturing of nanocomposites with embedded metal clusters showing the desired properties a good control of many experimental parameters, including properties of the substrate, as well as the plasma and discharge parameters is required. However, the high complexity of the underlying coupled processes makes a detailed understanding of the physics difficult. While there has been substantial progress in modeling processes in complex plasmas over the last years, e.g.\cite{springer-book,ReviewHenning} and powerful theoretical and computational tools have been developed in plasma physics and many-particle theory, e.g., \cite{bonitz03,rinton-book}, the coupling of plasma and surface processes is yet far too complex to allow a treatment based on microscopic approaches. The alternative is to apply simpler theoretical concepts and models which allow to perform simulations that are sufficiently large to capture at least some key features and trends of the experiment. The price one has to pay is that one has to use transport and rate coefficients as an input from more sophisticated simulations or from measurements. Here the method of choice are Kinetic Monte Carlo simulations (KMC), e.g. \cite{binder79}, which have seen remarkable progress over the last several decades and which have been successfully applied to study the behavior of metal adatoms on a metal surface, e.g., \cite{ziegler08} and references therein, and to crystal growth \cite{levi97} and cluster growth on surfaces \cite{thransimu,silverman}, for an overview see \cite{jensen99}. 

The aim of this paper is to present some key ideas of Kinetic Monte Carlo simulations and demonstrate how they can be applied to study deposition of metal atoms on a polymer substrate below the glass transition temperature and the formation of nanocomposites, for a general description of the KMC approach we refer to \cite{binder79}. Extending previous simulations \cite{thransimu}, here we present a more systematic analysis, including detailed results of the cluster size distribution and of the depth-resolved distribution of clusters inside the substrate. Special emphasis is devoted to an analysis of the dependence of the resulting material properties on the deposition rate of metal atoms. Finally, the effect of surface defects which tend to trap metal atoms is analyzed.

\begin{figure}[t]\label{fig:sketch}
\begin{center}
\includegraphics[width=0.8\linewidth]{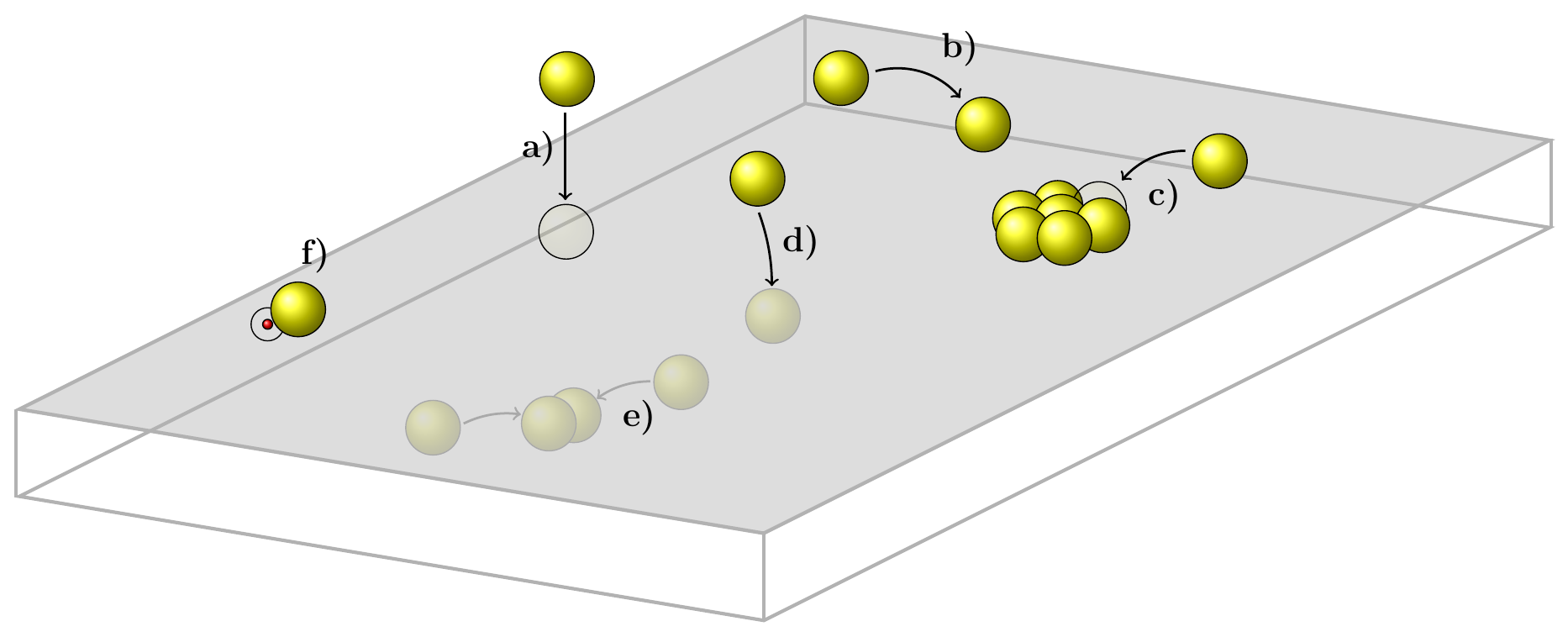}%
\end{center}
\caption{Main elementary processes considered in the simulations: a) deposition of atoms; b) surface diffusion of atoms and clusters\protect\footnotemark; c) cluster agglomeration on the surface; d) diffusion of clusters into the bulk; e) bulk diffusion and cluster growth in the bulk; f) trapping of clusters at defect sites. Other processes not shown and not studied in the present paper include the deposition of larger clusters and evaporation of clusters from the surface.}
\label{fig:sidecaption}
\end{figure}
\footnotetext{for simplicity atoms are also referred to as clusters from now on.}
\section{Description of the simulations}\label{simu_desc}
Atom and cluster deposition on a substrate in a plasma environment is a very complex process. Atoms arriving at the surface undergo a rather complex dynamics, some elementary processes are sketched in Fig.\,\ref{fig:sketch}: they include diffusion of clusters of varying size on the surface and into the bulk as well as merging and growth of clusters.
To gain a first understanding we have to make a number of simplifications which may be relaxed in future work. First, we will not consider cluster growth and charging effects in the plasma phase but will assume a given steady and homogeneous flow of atoms onto the surface. Further, no evaporation of atoms and clusters from the surface \cite{reemission} is included.
\subsection{The simulation model}
\label{model}
Our main goal is the analysis of the coupled diffusion and aggregation processes of metal atoms and clusters. Therefore, the complex microsopic properties of the polymer substrate, its atomic and chemical structure will not be directly included in the simulation. Instead, the main effect of the substrate on the clusters is condensed in effective average cluster mobilities which below will be contained in surface diffusion and bulk diffusion rates.

The substrate is located in the half-space $\{(x,y,z)\in \mathbf{R}^3\,|\,z\,\leq\,0\}$. To reduce finite size effects periodic boundary conditions in the $x$- and $y$-directions are applied.
Further, for clusters the simplest possible model is used. They are assumed to be spherical, and after aggregation of two clusters a spherical shape is assumed to be recovered sufficiently quickly. 
According to experimental results, the critical cluster size of nucleation is chosen to be one\cite{CriticalNucleus}.
The average density of the clusters is assumed to be constant regardless of their size and estimated to be equal to the bulk density of copper what means that a single atom is considered to have a radius $r_1$ of about $0.14\,\text{nm}$.
To model {\em diffusion} in a KMC scheme, the clusters are allowed to move along the surface without any preferred direction (isotropy) and into the bulk. A diffusion process consists of a jump of constant length $l=0.6\,\text{nm}$ which is chosen to be approximately the diameter of a polycarbonate chain. Jumps occur at discrete points in time with a probability (rate) depending on the material and cluster size, see below. Clusters on the surface $z=0$ may perform 
two different kinds of diffusion jumps: surface jumps and volume (bulk) jumps where the surface jump length equals the volume jump lengths.

The {\em nucleation of clusters} is mainly due to two processes. The first one is the formation of large clusters by agglomeration of two smaller clusters. 
The growth of clusters is assumed to be instantaneous, i.e. without any finite equlibration time. In the KMC simulations, two clusters agglomerate to a larger cluster if the distance of their surfaces falls below half of the jump length $l$. 
The second process that has to be taken into account is the growth of clusters at defect sites. To incorporate this effect, point-like defects are distributed randomly over the surface, they are assumed to have a trapping radius of $0.3\,\text{nm}$. If a cluster diffuses inside a circle with this radius it is trapped for the rest of the simulation and may serve as a nucleation site for large clusters. Despite its simplicity, the present model provides a balanced scheme for treating diffusion and cluster growth. While there may be uncertainties with respect to the resulting absolute time scales of the dynamics we expect that the main trends such as cluster size distribution, density profile or scaling with the deposition rate should be captured correctly. 

The {\em time evolution} of the system is governed by the rate coefficients $\nu^{n}_{b/s}$ (frequencies) of diffusion events which are related to the 
diffusion coefficients $\mathcal{D}^{n}_{s/b}$  by 
\begin{equation}
\label{diff_rate}
 \mathcal{D}^n_{b/s}=\beta_{b/s}l^2\nu^n_{b/s},\qquad \beta_{b/s}=\begin{cases} 1/6: \text{bulk diffusion}\\
                                                                                1/4: \text{surface diffusion} \end{cases},
\end{equation}
where $l$ is the jump length and $\beta $ a geometrical factor. Further, $b$ and $s$ stand for bulk and surface diffusion, respectively. The time scale in the whole simulation is then 
set by one of these rates. We use, as the basis for the time unit the (inverse of the) bulk diffusion rate of monomers
\begin{equation}
\tau = 1\,\text{jpa}:=1/\nu^1_b,
\end{equation}
which is the average time between two diffusion events of an atom (``jumps per atom'').
Typical values for the present system are a diffusion constant\cite{faupeldiffusion} $\mathcal{D}_b \sim 1.0\times10^{-13}\,\text{cm}^2/\text{s}$ and a jump length $l\sim 0.6\,\text{nm}$ leading to 
a time unit (1 jpa) of $\tau = 6.0\times10^{-3}\,\text{s}$.

{\em Size dependence of bulk diffusion.} Obviously, the diffusion coefficients depend on the cluster size which is indicated by the superscript $n$ denoting the number of atoms a cluster consists of. To achieve a reasonable model it is crucial to have an accurate picture of this dependence. For modeling bulk diffusion of clusters in polymer systems below the glass transition temperature different approaches were developed. A central role has played the so-called free volume theory (FVT) introduced by Cohen and Turnbull\cite{free_volume1} in the 1950s to describe the self-diffusion
in liquids of hard spheres. The main idea of FVT is to explain the diffusion in terms of the redistribution of vacancies. FVT predicts the following law for the bulk diffusion 
coefficient of the diffusing species
\begin{equation}
 \mathcal{D} \propto e^{-\gamma \frac{V^*}{V_f}},
\end{equation}
where $\gamma$ is a geometrical factor, $V^*$ is the minimum volume into which a molecule of the diffusing species can jump and $V_f$ is the average free volume per spherical molecule 
in the liquid. Cohen and Turnbull stated that $V^*$ is approximately the specific volume of the diffusing molecule what means that FVT, in its original formulation, predicts an
exponential decay of the diffusion rate with the size of the diffusing particle. The concepts of FVT were later extended to binary diffusion in systems consisting of
a polymer species and a solvent\cite{free_volume2}. These extensions also predict an exponential dependence of the diffusion coefficient on the particle size. Here we will follow 
free volume theory and use the following scaling of the bulk diffusion coefficient with the cluster size $n$
\begin{equation}
 \label{eq:decay_bulk_diff}
 \mathcal{D}^n_b=2^{-n}\cdot\mathcal{D}^{1}_b,\qquad\mathcal{D}^1_b\equiv1.
\end{equation}

{\em Surface diffusion.} To describe surface diffusion correctly the assumptions of the FVT are not adequate since the concept of vacancies is not appropriate. As a first approximation we use a simple power law to 
connect the surface diffusion constant of a cluster with its size
\begin{equation}
 \label{decay_surface_diff}
 \mathcal{D}^n_s=n^{-\alpha}\cdot\mathcal{D}^1_s,\qquad \mathcal{D}^1_s=m\cdot\mathcal{D}^1_b,
\end{equation}
where the monomer surface diffusion coefficient is given in units of the monomer bulk coefficient with a material-dependent coefficient $m$. We have varied both coefficients and checked the dependence of the results on them. For the simulation data presented below we used $\alpha = 1$ and $m=60$\cite{thransimu}. Molecular dynamics simulations of cluster diffusion on crystallline surfaces have also predicted a diffusivity following a power law of this form, where $\alpha$ is close to $1$\cite{jensen99}.

{\em Deposition rate and deposited density.}
In deposition experiments the metal atoms are deposited on the surface of the substrate during a finite deposition time $t_d$ with a certain flux of atoms $\mathcal{R}_d$,
which will be assumed stationary below. A common unit for $\mathcal{R}_d$ is the number of monolayers (ML) deposited during the time unit $\tau$.
In our simulations one ML corresponds to a surface density of 10 atoms per $\text{nm}^2$. Taking the same parameters as for the calculation of the timescale and assuming a thickness
of $0.3\,\text{nm}$ for one monolayer the unit 1ML/jpa corresponds to a physical deposition rate of $3.0\times10^{3}\,\text{nm/min}$ which is actually a very high deposition rate. 
In our simulations we varied the deposition rate by many orders of magnitude. The results presented below correspond to the range from $10^{-6}\dots 10^{-2}\,\text{ML/jpa}$. Assuming the same parameters
as above the simulated deposition rates translate into a range from $3.0\times10^{-3}\dots 3.0\times10^{1}\,\text{nm/min}$, that corresponds to typical experimental values\cite{faupeldepo}.

\subsection{Simulation algorithm}
The algorithm we applied to compute the time evolution of the described system goes back to a method that was introduced by Gillespie in the 1970s\cite{gillespie_algo1,gillespie_algo2} to simulate the time evolution of coupled chemical reactions. The main idea of the algorithm is to calculate time points for every possible transition or elementary process $i$ in the system (deposition or diffusion) with the associated rate $\nu_i$. These time points are stored in ascending order. The process corresponding to the smallest process time $t_{min}$ is carried out first and the system clock is advanced to the time point $t_{min}$. Next, the time points of processes belonging to those clusters that were involved
in that transition are removed from the list whereas for the remaining clusters a new process time $t_i$ is calculated and sorted into the list.

As the processes rates $\nu_i$ are assumed to be constant in time the probability distribution function $\mathcal{P}_i(t_i;t)$ for the time point $t_i$ of process $i$ is usually well described by a Poisson distribution 
\begin{equation}
 \mathcal{P}_i(t_i;t)=\nu_i e^{-\nu_i(t_i-t)}, \qquad \int_t^{\infty} dt_i \mathcal{P}_i(t_i;t) = 1,
\end{equation}
where $t$ is the actual time at which the time point $t_i$ ($t_i\ge t$) is being sampled.

\begin{figure}[h]
\includegraphics{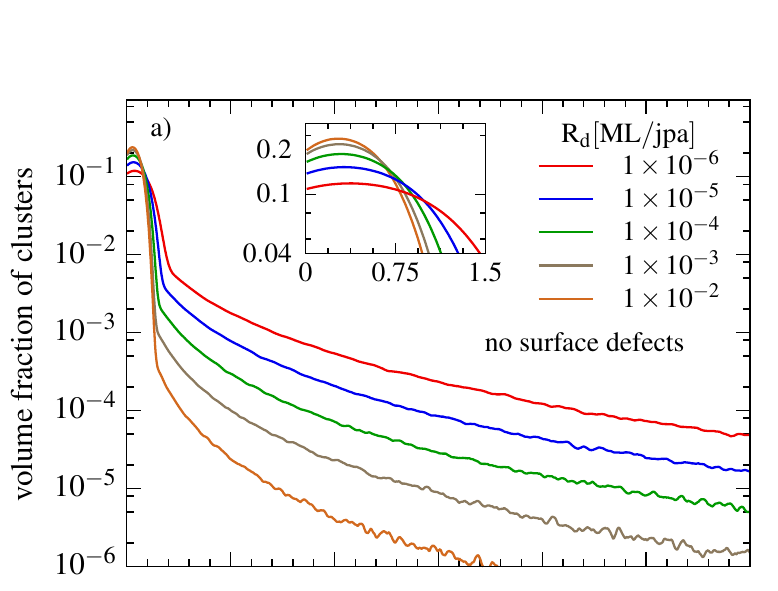}
\includegraphics{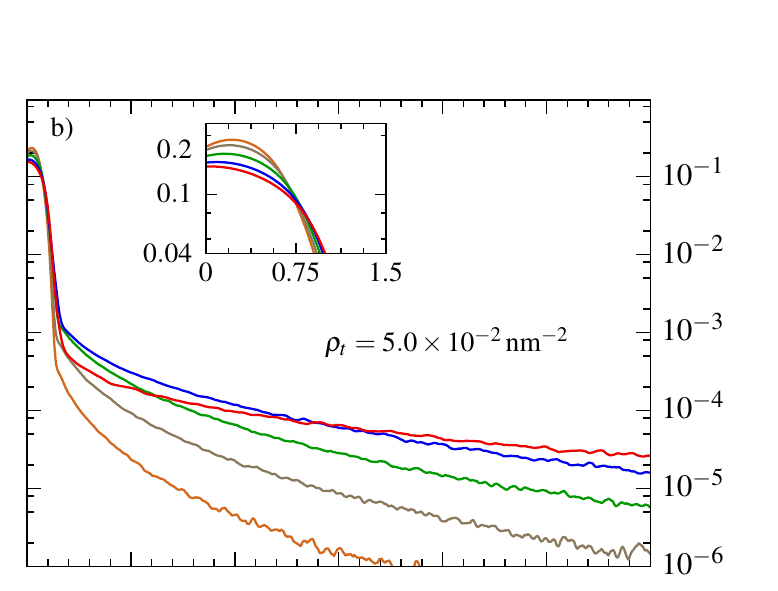} \\
\includegraphics{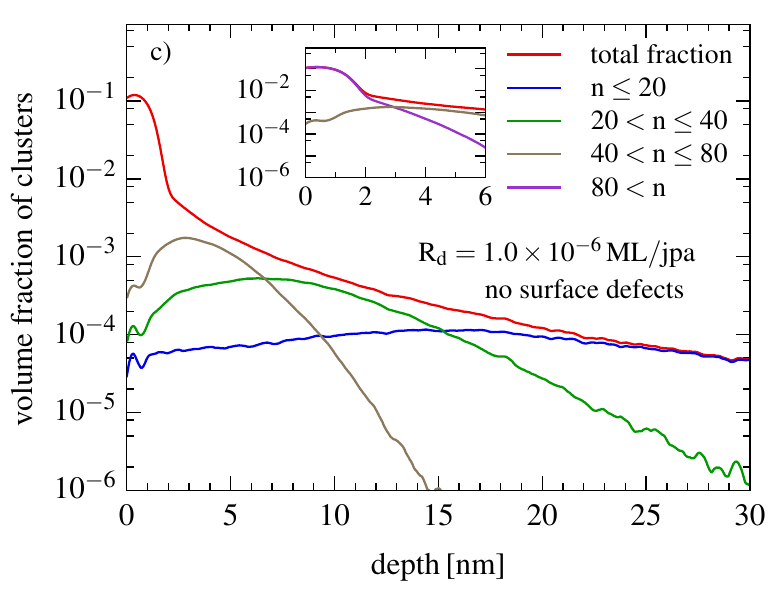}
\includegraphics{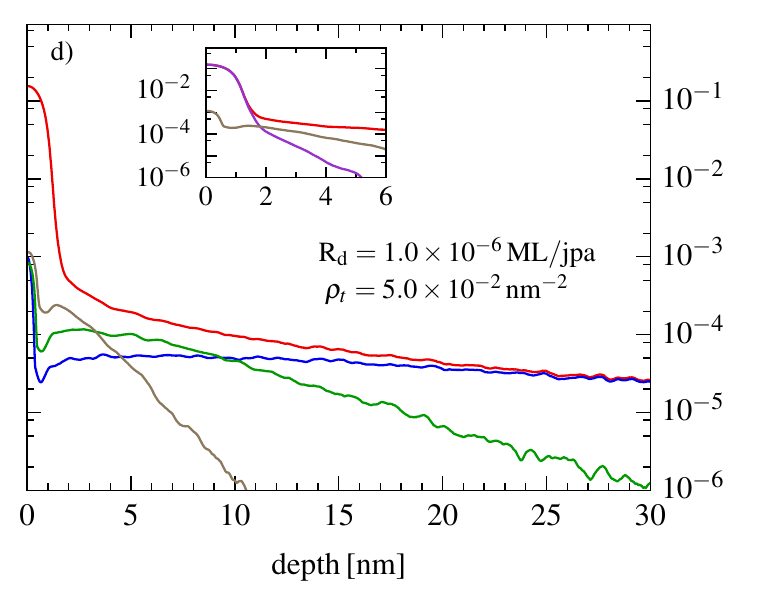}
\caption{Depth-resolved concentration profiles for five deposition rates indicated in the figure. Left (right) column: simulations without (with) surface defects.
Upper panel: total concentration profiles. Lower panel: contributions to the total concentration profile from clusters in different size ranges (different numbers of atoms $n$) indicated in the figure for the deposition rate $10^{-6}$ML/jpa. Insets show an enlarged view of the concentration profiles close to the surface. Note the semilogarithmic scale.}
\label{fig:conc}
\end{figure}
\section{Simulation results and discussion}\label{sec_results}
In this section we present the numerical results of our Kinetic Monte Carlo simulations. The main focus will lie on the influence of the density of surface traps and of the deposition rate on the system. For all results discussed below the size of the surface was chosen to be $2500\,\text{nm}\times2500\,\text{nm}$, the deposition rate was spatially homogeneous and varied between $1.0\times10^{-2}\,\text{ML/jpa}$ and $1.0\times10^{-6}\,\text{ML/jpa}$,
and the total amount of material deposited on the surface was set to 2 ML, corresponding to $6.25\times10^7$ atoms, for the given size of the surface. As in the experiments, metal atoms are deposited on the surface
over a final deposition time $t_d$ and immediately start to diffuse and agglomerate after deposition. All results presented below correspond to a time instant of $2.0\times10^5\,\text{jpa}$ after the termination of the deposition. We have carefully studied the time-dependence of our results and found that, after deposition of all atoms, no significant change of the results takes place. So the chosen time point is representative.
Further, simulations for an ideal surface (no defects) are compared to those with a fixed surface density of defects of $\rho_t=5.0\times10^{-2}\,\text{nm}^{-2}$. This is a value that is typical for ion bombardment experiments\cite{ionbomb,structural_modification}.

\subsection{Depth-resolved concentration profiles}\label{depth_resolved_concenration}
The first quantity we compute is the overall distribution of the adsorbate material inside the substrate. In Fig.\,\ref{fig:conc}a. and b. we plot the computed total concentration profiles (summed over all cluster sizes) as a function of substrate depth. Here ``concentration'' in a given (thin) layer at depth $z$ is defined as the volume occupied by clusters divided by the total volume of that layer.
The figures show concentration profiles for five different deposition rates ranging from $1.0\times10^{-6}\,\text{ML/jpa}$ to $1.0\times10^{-2}\,\text{ML/jpa}$. 

The total concentration profiles in Fig.\,\ref{fig:conc}a. can, in general, be divided into three parts: first, a narrow region just below the surface with a width of about $1$nm with a high clusters concentration, second a narrow region of a width between one and two nm near the surface where the concentration drops abruptly by one to three orders of magnitude and third, a broad region with a much slower concentration decrease into the bulk. The behavior in the third region is readily understood. Here the density decay is close to a Gaussian profile which would be the profile in the case of free diffusion of non-interacting particles according to Fick's second law (thin-film solution). Thus, all deviations from this ``normal'' decay are due to 
agglomeration and cluster growth. Cluster growth is strongest in the vicinity of the surface. Interestingly, the highest metal concentration is observed not at the surface but about one diffusion length below.

Let us now consider the effect of the deposition rate. With an increase of the rate, also the concentration in the surface region grows monotonically. However, deeper than about 1 nm the picture changes drastically: an increase of the deposition rate leads to a monotonic {\em reduction} of the metal fraction in the bulk. The explanation is that an increased deposition rate leads to a more homogeneous growth
of smaller, more or less immobile, clusters on the surface, which results in a higher coverage of the surface. Consequently the probability for small mobile clusters on the surface to encounter larger immobile clusters before diffusing into the bulk is enhanced. In addition to that, there is a higher probability for incoming atoms to impinge directly on clusters on the surface.
This explanation is readily verified by considering how many clusters of different sizes are found in a given depth. In Fig.\,\ref{fig:conc}c. the contributions of clusters with less than 20 atoms, with $20\dots 40$ atoms, $40\dots 80$ atoms and of larger clusters, to the total density are separated for one deposition rate. One clearly sees that, near the surface, the concentration is dominated by the largest clusters with a size of more than 80 atoms\,(see inset). This has been observed for all simulated deposition rates. 
Consider now the smaller cluster groups. For them we observe a maximum concentration in a depth of few nanometers: 
first, clusters with a size between 40 and 80 atoms reach a maximum concentration in a depth of about $3\,\text{nm}$, followed by clusters with $20\dots 40$ atoms the concentration of which peaks in about $7\,\text{nm}$ depth. Finally, clusters smaller than 20 atoms have their highest concentration in a depth of about $17\,\text{nm}$. Deeper in the bulk practically only the smallest clusters are found.


\subsection{Cluster size distributions}
\begin{figure}[h]
\includegraphics{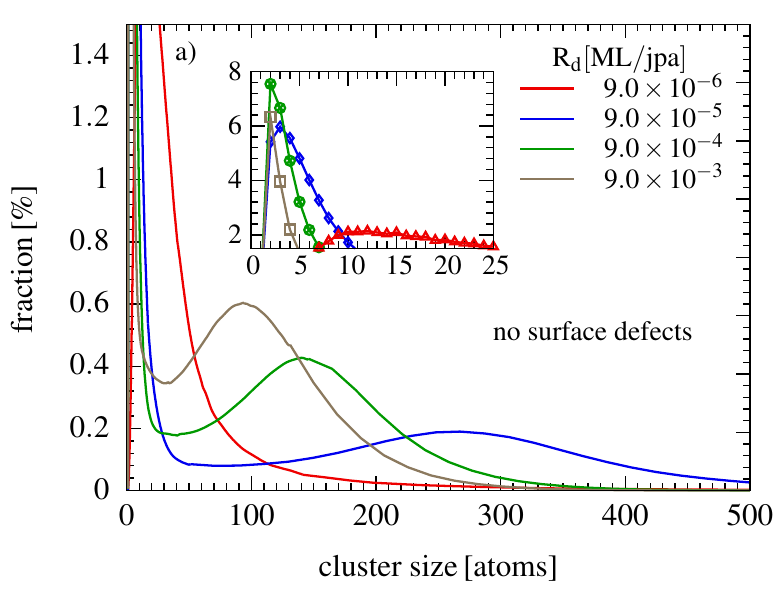}%
\includegraphics{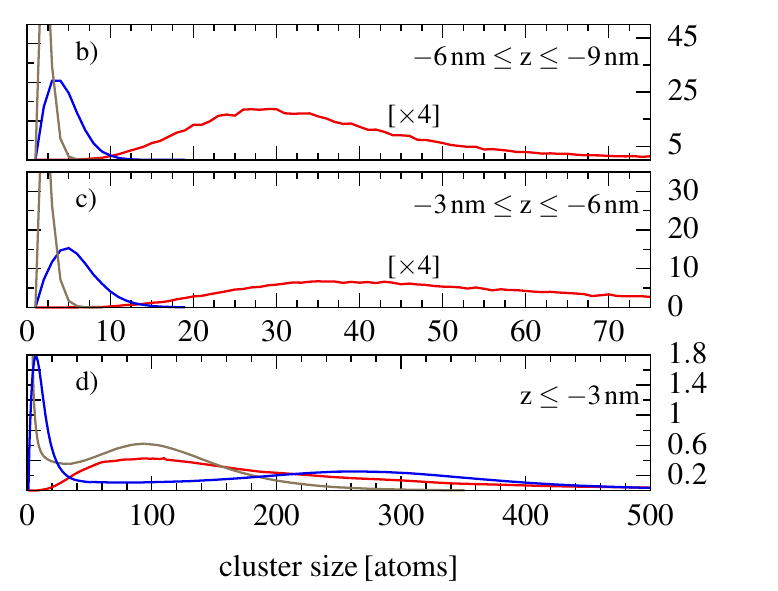}\\
\caption{Cluster size distributions for four different deposition rates shown in the figure. No surface defects.
{\bf Left}: global size distributions. {\bf Right}: depth-resolved size distributions in three different thin layers of thickness $3\,\text{nm}$ near the surface. Note the different vertical scales
            and the negative $z$-coordinate of the clusters.}
\label{fig:size1}
\end{figure}

In addition to the concentration profiles discussed above the simulations also yield the complete information about the size spectrum of the grown clusters. 
To this end, we calculated from the resulting cluster configuration the size distribution which is shown in Fig.\,\ref{fig:size1}. The left panel shows the 
global cluster size distributions for four different deposition rates. The size distributions typically have a bimodal
shape with two maxima which is also well known from transmission electron microscope measurements\cite{bib9}. The clear separation of the two peaks is remarkable 
although details vary with the deposition rate. An exception from this bimodal structure is observed for the lowest deposition rate (red curve) where only a single 
peak at small clusters with a size of around 10 atoms exists. However, a closer look shows that the behavior strongly varies with the depth, cf. Fig.\,\ref{fig:size1}.b-d:
There does exist a second (though weak) peak at larger cluster sizes, but its position is different in different layers. In the first 3 nanometers below the surface, the distribution peaks around $n=90$, in the next two layers the peak shifts to around $n=40$ and $n=30$, respectively. 

At all deposition rates except the smallest one the peak at small cluster sizes is by far the most dominant one. Thus there is a large fraction of dimers and trimers and a 
rapid decay of the size distribution towards larger clusters. It is interesting that an increase of the deposition rate leads to a sharpening of this first peak, see inset of Fig.\,\ref{fig:size1}.a. The analysis of our data shows that this sharp peak is related to cluster that are mainly located in the bulk. 

Consider now the second maximum. While its height is substantially below the one of the first peak, due to its large width, it contains a large total amount of atoms. This maximum at large cluster sizes results from the agglomeration of large clusters near the surface what is clearly confirmed by the right panel of Fig.\,\ref{fig:size1}. 
The shape and position of the second maximum at larger cluster sizes exhibits a strong dependence on the deposition rate.  

For high deposition rate\,(see the gray and green curve) the peak is located at smaller cluster sizes, between 100 and 150 atoms, and it is much narrower than for low deposition rates.
This reflects the tendency to homogeneous nucleation at the surface, in the case of high deposition rates. Furthermore, with decreasing deposition rate, the second maximum gets shifted towards larger cluster sizes, and it is clearly broadened (cf. blue curve).
This broadening results from an extension of the growth region of larger clusters near the surface\,(see also Fig.\,\ref{fig:conc}a) leading to a more heterogeneous nucleation.

\subsection{Dependence of the mean cluster size on the deposition rate}\label{ss:mean_size}
\begin{figure}[t!]
\includegraphics{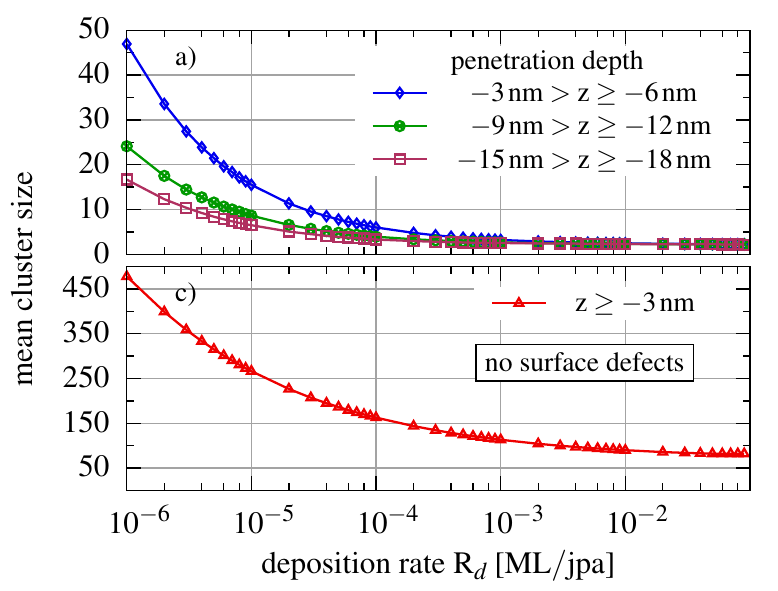}%
\includegraphics{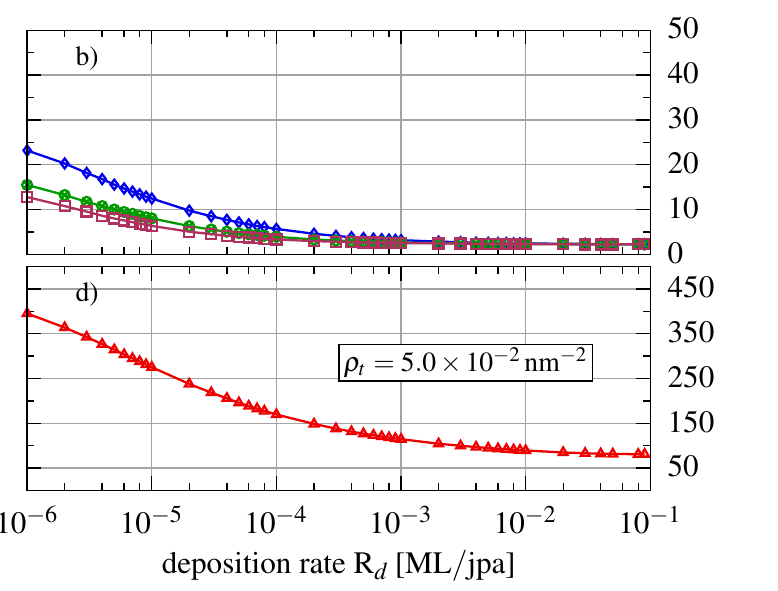}\\
\caption{Mean cluster size as a function of deposition rate. Bottom figures show the top layer, just below the surface, upper figures show three thin layers deeper in the bulk. Left (right) panel refers to an ideal surface without defects (a surface with a density of defects of $\rho_t=5.0\times10^{-2}\,\text{nm}^{-2}$).}
\label{fig:mean_size}
\end{figure}

In addition to the cluster size distribution we also analyze the mean cluster size and, in particular, how this average cluster size can be 
controlled by external parameters, most importantly, by the deposition rate. To answer this question we have performed a large number of additional simulations for a broader range of 
deposition rates than before including one order of magnitude lower and higher values and also more intermediate values. The results are collected in Fig.\,\ref{fig:mean_size}. The general trend obseved for any deposition rate is that 
the mean cluster size decreases with growing distance from the surface which is an obvious consequence of the scaling of the bulk diffusion coefficient with cluster size, Eq.\,(\ref{eq:decay_bulk_diff}). The decay is particularly strong immediately below the surface: from the first to the second layer the mean cluster size drops approximately by one 
order of magnitude. This is in agreement with the depth-resolved distribution of various cluster sizes presented in Fig.\,\ref{fig:conc}.c. Interestingly, below $9\,\text{nm}$, there is only marginal change of the mean cluster size at a given depostion rate. 

Now it is interesting to look at the absolute cluster size and its dependence on the deposition rate. These values are close to the peak positions of the cluster size distributions shown in Fig.\,\ref{fig:size1}.
It is impressive to note that, at a deposition rate of $10^{-6}\,\text{ML/jpa}$ very large clusters with a mean size close to $500$ atoms are formed at the surface, whereas in the layer below this size decays to approximately $\langle n\rangle = 50$.
Again there is a clear trend of a monotonic decrease of the mean cluster size with deposition rate which is observed in any depth. A saturation starts for deposition rates around $10^{-3}\,\text{ML/jpa}$. This decrease of the mean cluster size may, at first sight, seem surprising but it is consistent with our observations made in Fig.\,\ref{fig:size1}. The reason is that, at higher deposition rates, more small immobile clusters appear in the vicinity of the surface due to the homogeneous growth of clusters on the surface. As shown in Fig.\,\ref{fig:conc} in sec.\,\ref{depth_resolved_concenration} the enhanced
concentration of clusters at the surface inhibits the penetration of metal into the bulk, so with an increased deposition rate the contribution to the mean cluster size decreases in all depths.


\subsection{Influence of surface defects}\label{ss:defects}
We now turn to the analysis of nonideality effects on the substrate surface. There are many sources of defects which may result in fluctuations of the surface thickness or chemical composition. Here we concentrate on the influence of small point-like defects which act as trapping centers for the metal particles. It is also interesting to note that such defects may play a constructive role as they allow to reduce surface and bulk diffusion and to enhance nucleation at the surface, as it was shown in experimental work\cite{ionbomb}. Thus, it may be of interest to intentionally create such defects, e.g. by ion bomardment of the surface. Therefore, we have analyzed in detail the consequences of surface defects. Below we present results for a typical concentration of randomly arranged surface defects of
$\rho_t=5.0\times10^{-2}\,\text{nm}^{-2}$.

\begin{figure}[t!]
\includegraphics{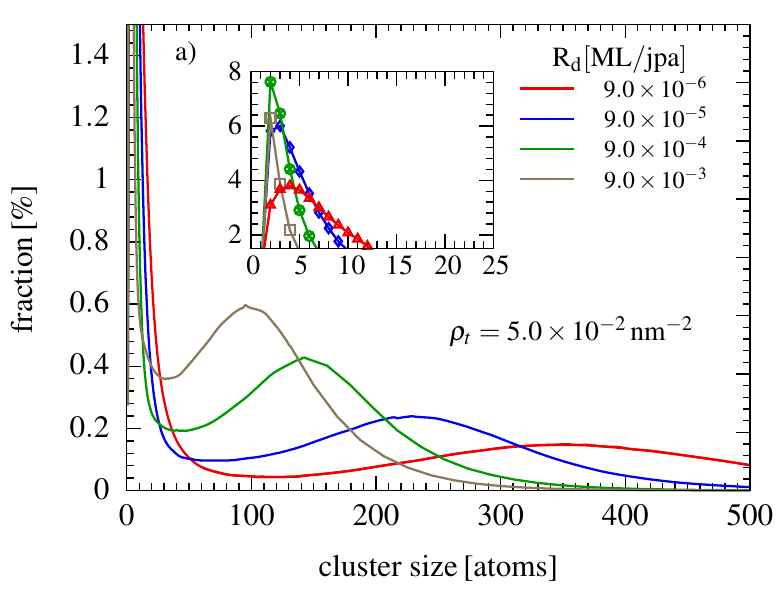}%
\includegraphics{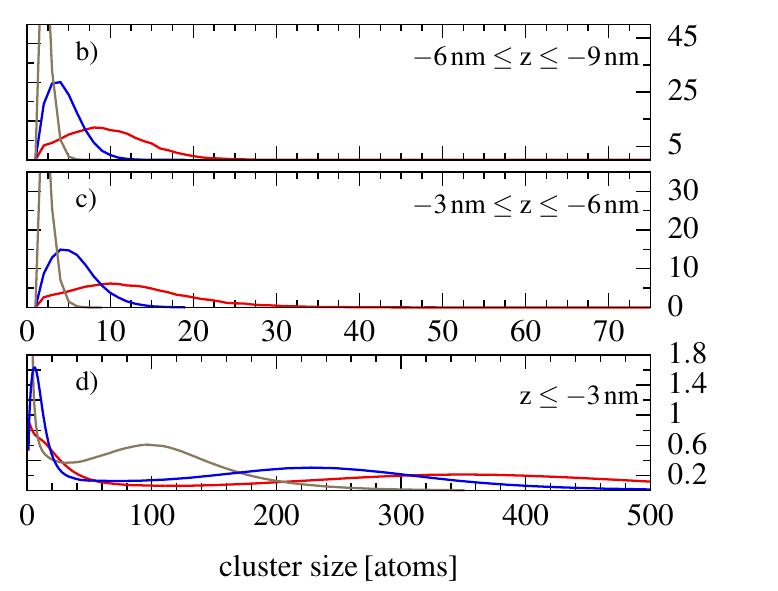}\\
\caption{Same as Fig.\,\ref{fig:size1}, but for the case of surface defects with a density of $\rho_t=5.0\times10^{-2}\,\text{nm}^{-2}$.}
\label{fig:size2}
\end{figure}

Let us start by looking again at the total depth-resolved concentration profiles, Fig.\,\ref{fig:conc}, and compare the previous results for an ideal surface with the case of surface defects included, Fig.\,\ref{fig:conc}b. From the figure it is obvious that the influence of surface defects strongly depends on the deposition rate: the shape of the concentration profiles is significantly influenced by surface defects only for the lowest deposition rate of $10^{-6}\,\text{ML/jpa}$. Here the concentration in the bulk is substantially lowered tending towards a flatter profile. At the same time, the concentration is enhanced in a narrow layer under the surface by almost a factor $2$. The obvious reason for this is the enhanced growth of clusters that are trapped at surface defects which act as nucleation centers. With increased deposition rate the nucleation on the surface is getting more homogeneous and the relative importance of the surface defects decreases.
This observation becomes even more evident by comparing the depth profiles of various cluster sizes in Fig.\,\ref{fig:conc}c and \ref{fig:conc}d. The red curve shows again the total concentration and the trends explained above. The other curves show the contribution of clusters in different ranges of size $n$. The defects have the strongest effect on the nucleation of the largest clusters, cf. brown curve refering to $n > 80$ atoms. The concentration of these clusters at the surface is one order of magnitude higher than without defects. For a depth larger than $2\,\text{nm}$ the concentration drops off much more rapidly than in the case without defects and practically vanishes for a depth exceeding $10\,\text{nm}$. Also, the maximum previously observed  in a depth of about $3\,\text{nm}$, is practically removed by the presence of defects.

Consider now the mean cluster size in the presence of defects, Fig.\,\ref{fig:mean_size} right panel. The overall trend is obvious: surface defects systematically reduce the mean cluster size.
This trend is particularly strong for deposition rates below $10^{-3}\,\text{ML/jpa}$ whereas for larger rates the influence of defects is negligible. Further, it is interesting to note that the influence of defects is felt differently in different depths. For example, for the lowest deposition rate defects reduce the mean cluster size at the surface by about $25\%$, but in the layers between $3\dots 12\,\text{nm}$ the cluster size is reduced by almost a factor of $2$. Deeper in the bulk, the influence of defects is again smaller. Thus surface defects, apparantly, provide a means to significantly modify the depth distribution of metal atoms.

Finally, we consider the effect of surface defects on the cluster size distributions which is presented in Fig.\,\ref{fig:size2} and is to be compared to the ideal case in Fig.\,\ref{fig:size1}. As seen before for the concentration profiles, surface defects have only a small effect for deposition rates exceeding $10^{-3}\,\text{ML/jpa}$. On the other hand, for lower rates, there is a noticeable effect on the size distribution, compare e.g. the red curves in the two figures\,($\text{R}_d=9.0\times10^{-3}\,\text{ML/jpa}$). Surface defects significantly narrow the first maximum and cause a shift to smaller cluster sizes, from $n\sim 10$ to $n\sim 4$. At the same time a second maximum (which was missing before at this deposition rate) appears at a mean cluster size of $n\sim 360$. This effect 
is particularly strong directly below the surface where the small peak observed previously around $n=100$ is now shifted towards $n\sim 350$, compare Fig.\,\ref{fig:size1}d and \ref{fig:size2}d. In contrast, in the layers between $3$ and $9\,\text{nm}$ the cluster size distributions have peaks around $n\sim 10$ whereas, without defects, a small peak in the range of $n=30\dots 40$ was observed.

\section{Conclusions and outlook}
In this paper we have presented a brief overview on Kinetic Monte Carlo simulations together with extensive results for metal deposition on a polymer substrate below the glass transition temperature. Owing to the simplicity of the model we were able to carry out simulations with a large number of atoms for a macroscopic piece of the surface with a dimension of $2.5\,\text{\textmu m}$ squared which allow to obtain results with a small statistical error. This was the basis for a systematic study of important trends related to the influence of the deposition rate. We have studied in detail how the metal atoms redistribute into the bulk of the substrate, how the cluster size distribution looks like and how clusters of different sizes are distributed inside the material. In agreement with earlier studies\cite{thransimu} and with experiments\cite{bib9} we find clear evidence for a pronounced bimodal size distribution which can be largely controlled by variation of the deposition rate. A high deposition rate gives rise to an almost monodisperse cluster distribution around dimers and trimers which propagate into the bulk. Furthermore, the second peak of the size distribution which is located near the surface can range from about $100$ to $400$ atoms, with a high deposition rate favoring smaller clusters. The presented model can be easily expanded to temperatures above the glass transition temperature, where the diffusion coefficients of large clusters are much higher due to the lowered viscosity of the polymer substrate and can be described by the Stokes-Einstein relation as a first approximation.

A second handle which may be used to control the spatial distribution of metal atoms is given by surface defects. In our simulations we used a random configuration of point-like defects which act as nucleation centers for clusters. While these defects are of minor importance in the case of high deposition rates, in the opposite case they have a profound influence on the resulting structure. We could show that the presence of defects substantially reduces the mean cluster size and that the strength of this effect is different in different depths. 
Such defects are easily produced experimentally by ion bombardment and could be useful to further optimize the structure of nanocomposites.

The present simulations are, obviously, based on a greatly simplified physical model. The main simplifications are the assumption of a spherical cluster shape formed by instantaneous agglomeration. For large deposition rates it is expected that non-spherical clusters form which may give rise to percolation. Work is in progress to remove these assumptions and to include a finite coalescence time based on a coalescence model described in\cite{coalescence1,coalescence2} and references therein.
A further simplification was the assumption of complete condensation of metal atoms on the polymer surface. In reality, metal atoms will have a finited probability of beeing reemitted\cite{reemission} after they spent some time at the surface. Since reemission is dominant for monomers though, we expect that omission of evaporation does not change the discussed quantities qualitatively but rather leads to a rescaling of the deposition rate. Nevertheless, it is of great interest to include the reemission of monomers and adjust the reemission rates with experimental data to compute 
instantaneous condensation coefficients\cite{InstantaneousCondensation}, which is defined as the ratio of the number of adsorbed atoms per unit time to the deposition rate.

From the view of metal deposition in a plasma environment, the next interesting questions to be included are the cluster growth in the gas phase giving rise to deposition of a pre-shaped cluster distribution. 
Also, spatial inhomogeneities, charging effects and time-resolved effects might be of interest and will be included in our model in future work.

\section*{Acknowledgements}
This work is supported by the Deutsche Forschungsgemeinschaft via the SFB-TR24, projects A7 and B13.


\end{document}